\documentclass[]{mn2e}
\usepackage{psfig}
\usepackage{epsf}
\newcommand{\ltaraw}{$\; \buildrel < \over \sim \;$}
\newcommand{\lta}{\lower.5ex\hbox{\ltaraw}}
\newcommand{\gtaraw}{$\; \buildrel > \over \sim \;$}
\newcommand{\gta}{\lower.5ex\hbox{\gtaraw}}

\loadboldmathitalic 
\title [Resolved nuclear CO(1-0) emission in APM08279+5255]
{Resolved nuclear CO(1-0) emission in APM08279+5255: \\ 
Gravitational lensing by a naked cusp?} 
\author[G. F. Lewis et al.]
{Geraint F. Lewis$^{1}$, 
Chris Carilli$^{2}$, 
Padeli Papadopoulos$^{3,4}$ \&
R. J. Ivison$^{5}$ \\
$^{1}$
Anglo-Australian Observatory, P.O. Box 296, Epping, NSW 1710, Australia:
Email \tt{gfl@aaoepp.aao.gov.au}\\
$^{2}$
National Radio Astronomy Observatories , PO Box 0, Socorro, New Mexico
87801-0387, USA:
Email \tt{ccarilli@nrao.edu}\\
$^{3}$
Astrophysics Division, Space Science Dept. of ESA, ESTEC, Postbus 299, NL-2200
AG, Noordwijk, The Netherlands  \\
$^{4}$
Sterrewacht Leiden, P.O. Box 9513, 2300 RA Leiden, The Netherlands:
Email \tt{ppapadop@astro.estec.esa.nl}\\
$^{5}$
Astronomy Technology Centre, Royal Observatory, Edinburgh,  
Blackford Hill, Edinburgh, EH9 3HJ, UK:
Email \tt{rji@roe.ac.uk}\\
}
\date{\today}
\begin{document} 
\maketitle 
\begin{abstract}
The ultraluminous broad absorption line quasar APM08279+5255 is one of
the most luminous systems known.   Here, we present an analysis of its
nuclear CO(1-0) emission. Its  extended distribution suggests that the
gravitational  lens in this  system is  highly elliptical,  probably a
highly inclined disk.  The quasar  core, however, lies in the vicinity
of  naked  cusp,  indicating  that  APM08279+5255 is  truly  the  only
odd-image gravitational  lens.  This source  is the second  system for
which the gravitational lens can be used to study structure on sub-kpc
scales in the molecular gas  associated with the AGN host galaxy.  The
observations and lens model require CO distributed on a scale of $\sim
400$ pc.  Using this scale, we  find that the molecular gas mass makes
a significant,  and perhaps dominant,  contribution to the  total mass
within a couple hundred parsecs of the nucleus of APM08279+5255.
\end{abstract}

\newcommand{\hawaii}{Hawaii~167}
\newcommand{\iras}{FSC~10214+4724}
\newcommand{\apm}{APM08279+5255}

\section{Introduction}\label{introduction}
Identified serendipitously in a search for high latitude carbon stars,
the $z=3.9$ broad  absorption line quasar \apm\ is  coincident with an
IRAS source with  a flux of 0.95Jy at 100$\mu$m  (Irwin et al.  1998).
Observations  with SCUBA  reveal  that \apm\  possesses a  significant
submillimetre  flux  of  75mJy  at  850$\mu$m (Lewis  et  al.   1998),
implying  a   bolometric  luminosity  of  $\sim5\times10^{15}L_\odot$.
Imaging reveals that  \apm\ is not point-like, but  rather is extended
over  a  fraction of  an  arcsecond  with  a structure  indicative  of
gravitational lensing  (Irwin et al.  1998).  The  composite nature of
\apm\ was confirmed in adaptive  optics (AO) images obtained by Ledoux
et  al.  (1998), with  the system  appearing as  a pair  of point-like
images  separated by  0.4  arcsec.  Observations  with  NICMOS on  the
Hubble Space Telescope  (Ibata et al.  1999) and  AO images taken with
the Keck telescope (Egami et  al. 2000) also uncovered a fainter third
component between the brighter two.  Gravitational lens models derived
from these  observations suggest that the quasar  continuum source has
been magnified by $\sim90$.

Using IRAM,  Downes et  al.  (1999) detected  emission in  CO(4-3) and
CO(9-8),  revealing the presence  of warm  circumnuclear gas  in \apm.
Papadopoulos  et al.   (2001)  were  able to  search  for CO(1-0)  and
CO(2-1) in  this system using the  Very Large Array  (VLA).  Both were
clearly detected associated with the quasar nucleus, as well as a more
extended  component located  several arcsecs  from the  quasar images.
Using locally  established values of the CO-to-H$_2$  ratio, this lone
cloud  represents  $\sim10^{11}M_\odot$  of cold  and/or  subthermally
excited gas.

In this paper,  we present an analysis of  nuclear CO(1-0) emission in
\apm\ using  VLA~\footnote{The VLA is  operated by the  National Radio
Astronomy  Observatory, which is  a facility  of the  National Science
Foundation,  operated   under  cooperative  agreement   by  Associated
Universities, Inc.}  at high  spatial resolution (0.3 arcsec).  The CO
appears as  a partial ring  of $\sim$0.6 arcsec diameter.   These data
suggest  a total  revision in  the gravitational  lens model  for this
source, with the  new model involving a `naked  cusp', which naturally
accounts for the observed odd-number  of images.  They also imply that
the nuclear CO  must be spatially extended on a scale  of at least 400
pc, making this  the second source in which  gravitational lensing can
be  used  as a  `telescope'  to  explore  sub-kpc scale  structure  of
molecular gas in the AGN host galaxy.

\begin{figure*}
\centerline{ \psfig{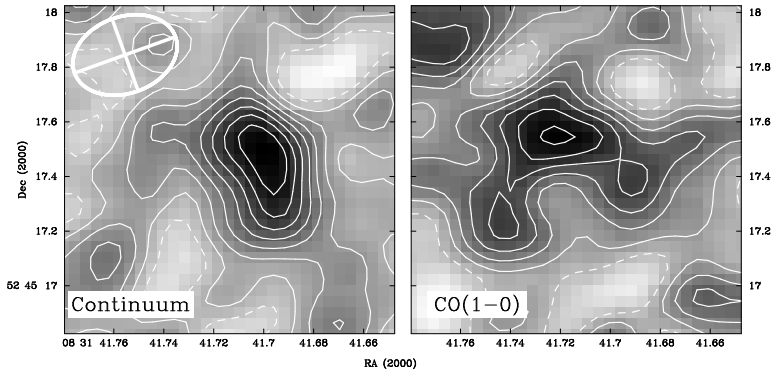}}
\caption[]{The  23Ghz  continuum  (left)  and  CO(1-0)  line  emission
(right)  in \apm.  The  contours are  at  -1 (dashed),0,1...   $\times
\sigma$.   The image has  been CLEANed  and the  ellipse in  the upper
corner of the left-hand panel represents the CLEAN beam.  The quasar A
image, as determined  from NICMOS imaging, lies at 08  31 41.64, 52 45
17.5, offset  from the  radio emission (Ibata  et al. 1999).   This is
probably  due to  astrometric  errors  in the  VLA  and HST  reference
frames.}
\label{fig1}
\end{figure*}

\section{Observations and Results}\label{observations}
Observations of the CO(1-0) emission from \apm\ were made in March and
April 2001 using  the VLA in the B (10 km)  configuration.  A total of
20 hours  were spent on the  source.  Due to limitations  with the VLA
correlator, we  chose to  observe in continuum  mode using two  50 MHz
bandwidth IFs, each with two polarizations. One IF was centered on the
redshifted CO(1-0)  line, corresponding  to a frequency  of 23.465GHz,
while  the second  IF  was tuned  away  from the  emission  line to  a
frequency  of 23.365GHz. This  observing set-up  maximizes sensitivity
since the effective  bandwidth (45 MHz, 575 km/s)  closely matches the
obseved  CO linewidth (Downes  et al.  1999), but  sacrifices velocity
information (Carilli, Menten, \& Yun 1999).  The source 3C286 was used
for absolute  gain calibration.  The rms  noise on the  final image is
35$\mu$Jy  beam$^{-1}$   with  an  effective   spatial  resolution  of
FWHM=$0.39''\times  0.28''$  with  a  major  axis  position  angle  of
$-70^o$.   Fast  switching  phase  calibration  was  employed  with  a
calibration duty cycle of 130  seconds (Carilli \& Holdaway 1999). The
phase stability for all the  observations was excellent, such that the
array was easily phase coherent  over the calibration cycle time. This
was  demonstrated by imaging  the phase  calibrator (0824+558)  with a
similar  calibration cycle time,  and by  imaging the  radio continuum
emission from \apm\ itself.

The images of the line and  continuum emission from \apm\ are shown in
Figure  1. The CO  line image  was generated  by subtracting  from the
visibility data a Clean-Component model of the continuum emission made
from the off-line data.  The peak surface brightness for the continuum
emission is $0.24\pm0.035$ mJy  beam$^{-1}$, with a total flux density
$0.41\pm0.07$ mJy.   The corresponding  numbers for the  line emission
are $0.183\pm0.035$ mJy beam$^{-1}$  and $0.39\pm0.09$ mJy. The latter
corresponds  to a  velocity-integrated line  flux of  $0.22\pm0.05$ Jy
km/sec, consistent with $0.15\pm0.045$ Jy km/sec deduced for the inner
$\sim1$arcsec by Papadopoulos et al. (2001)

The continuum emission is  extended north-south, with a position angle
and spatial  extent as expected  based on the three  component optical
gravitational lens. An attempt was made to decompose the continuum map
into three point-like images,  centred upon the positions derived from
the  analysis of  HST images  (Ibata et  al. 1999).   Given  the image
resolution,  it was not  possible to  cleanly separate  the north-most
images  (A+C) which  have  a separation  of  $\sim$0.15 arcsecs.   The
relative flux of (A+C)/B$\sim$1.5,  quite similar to the optical ratio
presented by  Ibata et  al. (2001).  The  line emission, on  the other
hand, shows distinct curvature away from the continuum position angle,
and at the 3$\sigma$ surface  brightness level it appears as an almost
complete ring with  a diameter of about 0.6 arcsec.   While the map of
the line emission  is of low signal to  noise, the ring-like structure
is apparent in each of  several days worth of observations.  Also, its
size is  in agreement  with that of  the CO(4-3) and  CO(9-8) emission
(Downes  et  al.  1999)  and  the  CO(2-1)  emission (Papadopoulos  et
al. 2001). Hence, these observations  reveal the gross features of the
CO(1-0) emission,  although more observations are  required to uncover
the finer details.

The observed CO(1-0) brightness temperature of the ring (averaged over
$\sim$575 km/sec)  is 1.4$\pm$0.8 K,  which corresponds to  an emitted
brightness temperature  of $T_b\sim$7 K at z$\sim$3.9.   A lower limit
for the  magnification factor of  the CO(1-0) emission can  be derived
assuming  the  warm  gas  emitting  the high-J  CO  lines  (Downes  et
al. 1999) to be also emitting  the J=1-0.  This gas phase is optically
thick  with $T_{kin}\sim200K$,  in  agreement with  the inferred  dust
temperatures   (Lewis   et   al.    1998),   which   then   yields   a
velocity-averaged  filling   factor  of  $f\sim7/200=0.035$.   Since
differential  lensing will  ``boost''  a compact  warm  region at  the
expense  of a  more extended  and possibly  sub-thermally  excited gas
phase that emits the CO J=1-0, the true value of f will be larger.

\section{Gravitational Lensing}\label{lensing}
The lensing galaxy  has yet to be identified in  \apm. Hence, only the
quasar positions and magnitudes are available to constrain any lensing
model.  Other than \apm, all other lens systems possess an even-number
of images.   Theoretically, any non-singular  mass distribution should
produce an odd number of  lensed images (Burke 1981), and the ubiquity
of even  numbers of images has been  used to limit the  core radius in
lensing systems, as  small cores result in the  demagnification of one
of the images.  Due to  the brightness of the central source, however,
models for  \apm\ have  required the opposite,  a very  circular model
with a  large core /  shallow cusp (Lewis  et al.  1999; Egami  et al.
2000; Munoz et al.  2001).  We further explore the lens model of \apm\
in light of the observations presented here.

\begin{figure*}
\centerline{ \psfig{figure=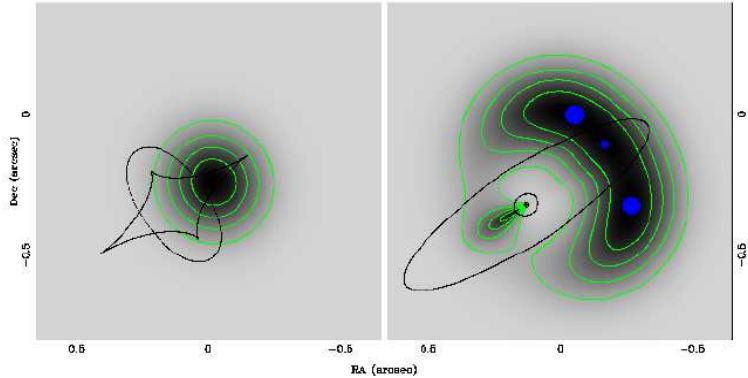,angle=270,width=4.0in}}
\caption[]{The  source plane (left)  and image  plane (right)  for the
gravitational  lens  model  for  \apm.   The  diamond  and  elliptical
caustics, and their corresponding  critical lines, are apparent in the
panels. The  contour lines represent the  CO source and  image and the
filled circles represent the  positions and brightnesses of the quasar
images as  determined from the HST  images (Ibata et  al. 2000). Note,
however, the model presented in the paper results in brightness ratios
of $C\sim B\sim0.75A$, whereas the observed ratios are $C\sim0.2A$ and
$B\sim0.8A$.}
\label{fig2}
\end{figure*}

The  concordance between  the optical  images of  \apm\ and  the radio
continuum at  3.6cm (Ibata  et al. 1999)  and 23GHz  demonstrates they
come  from coincident regions  with a  similar scale  size. Associated
with  the active  core of  \apm, these  regions are  smaller  than the
gravitational  lensing  caustics.    The  quite  different  morphology
displayed  in the  CO  image, however,  indicates  that this  emission
arises  in a  larger region,  distinct from  the  continuum radiation.
With  its  extended nature,  the  CO  emitting  region can  lie  under
different parts  of the caustic  network. The CO source,  however, can
not be  significantly more extended  than the caustic network,  as the
resulting   image  would  not   show  the   ring  structure   seen  in
Figure~\ref{fig1}.  Therefore, for a given model, the size of the ring
image provides a probe of the scale size of the emitting region.

To  form ring-like images,  the source  must be  extended and  cover a
substantial fraction of  the inner caustic structure of  the lens; the
resulting image  is appears  as a ring,  following the  outer critical
line  in  the image  plane  (see  Kochanek,  Keeton \&  McLeod  2001).
Examining  the critical  line  structure in  all previously  published
models for  \apm\ (Ibata et al. 1999;  Egami et al. 2000;  Munoz et al
2001),  the resulting  structure for  the  Einstein ring  of a  source
centred  upon the  quasar nucleus  should  be quite  circular, with  a
radius of  $\sim0.2$ arcsecs, passing through the  two brighter quasar
images; such images  for extended source can be seen  in the models of
Egami  et al. (2000).   This is  quite different  to the  CO structure
displayed in Figure~\ref{fig1}, with the CO emission clearly showing a
roughly  east-west extension,  with  the hole  in  the ring  occurring
$\sim0.5$ arcsecs from image A.   Using the model derived from the HST
data  (Ibata et  al.  1999),  we explored  the image  configurations a
range of sizes and shapes, centred  upon the quasar source, for the CO
emitting region.  None reproduced  the observed image structure and we
reject this previous model.
 
Naked  cusps occur  in highly  elliptical systems,  such  as flattened
disks, when  the inner diamond caustic extends  outside the elliptical
caustic.   A  source  inside  this extension  produces  three  roughly
colinear images of similar brightness (Maller, Flores \& Primack 1997;
Bartelmann  \& Loeb  1998; Keeton  \& Kochanek  1998; Moller  \& Blain
1998;  Blain,   Moller  \&  Maller  1999;   Bartelmann  2000).   While
gravitational  lensing  by spiral  galaxies  has  been observed  (e.g.
B~1600+434; Koopmans, Bruyn \& Jackson 1998) no observed lensed quasar
system has so far been associated with a naked cusp.

To examine  the possibility that  the observed image  configuration is
due to gravitational lensing by  a naked cusp, we constructed a simple
mass model.  Following  Batelmann and Loeb (1998), this  consists of a
truncated  flattened disk in  a spherical  halo. \apm\  has brightened
considerably since its  discovery (Lewis, Robb \& Ibata  1999; E. Ofek
priv.   comm.,   see  {\tt  http://wise-obs.tau.ac.il/$\sim$eran/LM}),
potentially  due   to  the  effects   of  gravitational  microlensing.
Comparing the images obtained in 1998 (Ibata et al. 1999; Egami et al.
2000) and those  obtained in 1999  (Munoz et al.  2001),  the relative
image brightness have changed  appreciably, with image B changing from
being 78\% of image A to only 50\% in the latter epoch; such behaviour
is extremely  suggestive of the action  of gravitational microlensing,
although longer  term monitoring is  required to confirm  this. Hence,
the relative image  brightnesses cannot be used to  constrain any mass
model. Given  the sparsity of constraints,  we choose to  find a model
that  can recover the  image positions,  while providing  a reasonable
description  of the  observed CO  emission.  As  can be  imagined, the
parameter space  is large,  so a range  of models that  reproduced the
quasar positions were chosen and  then modeling of the CO emission was
undertaken  by-eye. With  this, therefore,  we do  not claim  that the
model presented here is unique,  only consistent with the general form
of data.

In  our  chosen model,  the  disk  is  highly inclined,  presenting  a
projected   axis  ratio   of   0.25,  and   the   disk  truncated   at
~8$h^{-1}$kpc~\footnote{$\Omega=1$   and    $\Lambda=0$   is   assumed
throughout},  and  possess a  core  radius  of 0.065$h^{-1}$kpc.   The
rotation  velocity of  the disk  is  200km$s^{-1}$.  Figure~\ref{fig2}
presents  the  source  and  image  plane  for  this  model,  with  the
elliptical  and  diamond  caustic,  and corresponding  critical  lines
apparent.   In  this model,  the  quasar  images  are not  substantial
magnified, with  a total magnification of $\sim7$,  with the intrinsic
source      of     \apm\      being      correspondingly     luminous,
$L_{bol}\sim7\times10^{14}L_\odot$. While  extreme, this value  is not
necessarily outrageous  as the unlensed quasar  HD1946+7658 possess an
intrinsic   luminosity   of   $\sim4\times10^{14}L_\odot$  (Hagan   et
al. 1992),  and \apm\ may be a  member of this very  luminous class of
quasars.  It must be conceded, however, that the non-uniqueness of the
lens model translates into  uncertainty in the model magnification and
a  true determination  of the  intrinsic properties  of  \apm\ require
models derived from better observational constraints.

The  CO source  is taken  to be  have an  circular  surface brightness
distribution centred  upon the quasar position.   One important aspect
of  the results  presented herein  is  that \apm\  becomes the  second
system for which the gravitational lens can be used to study structure
on sub-kpc  scales in the molecular  gas associated with  the AGN host
galaxy, the first system being  the Clover Leaf quasar, H1413+117 (Yun
et  al.  1998;  Kneib, Alloin,  \& Pello,  R.  1998).   For  \apm, the
observations  and lens model  require the  CO to  be distributed  on a
scale covering  a substantial  fraction of the  caustics in  the image
plane, but not  too large to lose the ring  structure. The lower limit
to the  CO source  size based on  the modeling is  $\sim400h^{-1}$ pc,
while a rough upper limit is $\sim$1 kpc. With this model, the CO(1-0)
has been magnified by a  factor of $\sim2.5-3$.  Like the Clover Leaf,
we find that the spatial extent and mass of the molecular gas in \apm\
are  comparable to  those seen  in nearby  nuclear  starburst galaxies
(Sanders and Mirabel 1996; Downes and Solomon 1998).

Downes   et   al.    (1999)   determined   a   CO   source   size   of
$\sim(80-135)h^{-1}$pc  for  the  estimated magnification  factors  of
$\sim20-7$. This size  is much smaller than the  one calculated above,
while their magnification factors are larger.  Their analysis is based
upon   modeling  of   CO  emission   in  the   gravitationally  lensed
ultraluminous  galaxy  IRAS F10214+4724  (Downes,  Solomon \&  Radford
1995),  whose image  is  clearly an  extended  arc-like feature  which
possesses an essentially linear  magnification. Such a simple model is
probably a  poor representation of the lensing  in \apm. Additionally,
Downes  et al.  (1999) assumed  that  the velocity  filling factor  is
unity,    substantially   larger   than    the   value    derived   in
Section~\ref{observations};  as the intrinsic  source radius  in their
model scales inversely with this value and the magnification factor is
proportional to it. For f$\sim$0.35, a value well within our estimated
range, the  Downes et al. (1999)  model yields an upper  limit for the
intrinsic source size  and a lower limit for  the magnification factor
that our similar to ours.

Accounting   for   the  influence   of   gravitational  lensing,   the
velocity-integrated  CO(1-0)  flux density  implies  that the  nuclear
content  of molecular gas  in \apm\  is $\sim10^{10}  h^{-2} M_\odot$,
assuming a CO-to-H$_2$ conversion  factor of ${\rm\sim 1 (M_\odot\ km\
sec^{-1}\  pc^2)^{-1}}$ which  is typical  for starburst/kinematically
violent, UV-intense environments of gas-rich, IR-ultraluminous systems
(Downes \& Solomon 1998).  Assuming the  CO is in a rotating disk, the
dynamical mass can  be calculated from the radius  of $\sim500$ pc set
by  the lens  modeling,  and using  a  rotational velocity  of 350  km
sec$^{-1}$  set  by  the observed  of  line  velocity  HWHM =  250  km
sec$^{-1}$ and assuming a disk  inclination angle of 45$^o$ (Downes et
al.   1999).   The  implied  dynamical  mass  is  $1.5\times  10^{10}$
M$_\odot$  within $\sim$500  pc of  the nucleus,  consistent  with the
value  derived from the  CO flux.   Hence it  appears likely  that the
molecular  gas  mass  makes   a  significant,  and  perhaps  dominant,
contribution to  the total  mass within a  few hundred parsecs  of the
nucleus in  \apm, unless the nuclear  CO disk is close  to face-on.  A
similar conclusion has been  reached for most nearby nuclear starburst
galaxies (Downes and Solomon 1998).

\section{Conclusions}\label{conclusions}
This paper  has presented resolved images of  nuclear CO(1-0) emission
in  the gravitationally lensed  BAL quasar  \apm. While  the continuum
emission is found  to be well aligned with  the optical quasar images,
the CO(1-0) is more extended, with a broken ring-like appearance. Such
a structure  is consistent with  the action of  gravitational lensing,
with the continuum emission occurring on the scale of the quasar core,
while the  CO(1-0) arises from  a larger region and  is differentially
magnified. The  three-image nature  of \apm\ has  posed a  problem for
lens modeling, as an extremely large, flat core is required to produce
the  central image.   Such  three image  configurations  are a  nature
consequence of  gravitational lensing  by a flattened  potential which
can produce  naked cusps.  Modeling  of the CO(1-0)  emission supports
this hypothesis,  although a deficit  in constraints implies  that the
model is not unique. An immediate prediction of this model is that the
lensing galaxy,  whose position could  be revealed by  observing below
the Lyman limit for this system ${\rm (\lta 4400\AA)}$, hence removing
the glare from the quasars,  should be offset $\sim0.5$arcsec from the
quasar image, rather than lying behind the quasar images.

Currently,    our    CO   images    of    \apm\    are   of    limited
signal-to-noise. However,  with further integration a  detailed map of
the CO image can be made. As this region will be free from the effects
of  microlensing,  and  as  its  extended nature  provides  many  more
constraints (Kochanek,  Keeton \& McLoed  2001), such imaging  has the
potential  to provide a more  accurate model  of the  lensing in
\apm\ than from the quasar images.

\end{document}